\documentclass[aps,prl,reprint,superscriptaddress]{revtex4-2}

\usepackage{graphicx}
\usepackage{amsmath}
\usepackage{bm}
\usepackage{physics}
\usepackage{hyperref}
\usepackage{xspace}
\usepackage{xcolor}

\newcommand{\trento}{T\raisebox{-0.5ex}{R}ENTo}

\begin{document}

\title{Tracing Electromagnetic Response through Charge Splitting of Directed Flow in the Quark–Gluon Plasma}

\author{Ankit Kumar Panda}
\affiliation{Key Laboratory of Quark and Lepton Physics (MOE) \& Institute of Particle Physics,
Central China Normal University, Wuhan 430079, China}

\author{Zhu Meng}
\affiliation{Key Laboratory of Quark and Lepton Physics (MOE) \& Institute of Particle Physics,
Central China Normal University, Wuhan 430079, China}

\author{Long-Gang Pang}
\email[]{lgpang@ccnu.edu.cn}
\affiliation{Key Laboratory of Quark and Lepton Physics (MOE) \& Institute of Particle Physics,
Central China Normal University, Wuhan 430079, China}

\author{Xin-Nian Wang}
\email[]{xnwang@ccnu.edu.cn}
\affiliation{Key Laboratory of Quark and Lepton Physics (MOE) \& Institute of Particle Physics,
Central China Normal University, Wuhan 430079, China}

\begin{abstract}
Experimentally observed charge splitting of directed flow, \(\Delta v_1\equiv v_1(h^+)-v_1(h^-)\), exhibits both a sign reversal at midrapidity and the amplitude change from Au+Au collisions at \(\sqrt{s_{\rm NN}}=200\) GeV to Pb+Pb at \(\sqrt{s_{\rm NN}}=5.02\) TeV. We study this behavior within a unified \((3+1)\)-dimensional viscous hydrodynamic model that incorporates an electromagnetic source term in the bulk evolution of the quark-gluon plasma (QGP) and an electromagnetic correction to the Cooper--Frye particlization. The beam-energy dependence of $\Delta v_1$ can be described by varying the effective electrical conductivity between RHIC and LHC energies, with values consistent with the conductivity scale indicated by lattice-QCD calculations. We find that the sign reversal emerges from the competition between momentum distortions induced by the electric and magnetic field which is controlled by the conductivity during the QGP evolution. The model further predicts the mass-dependent charge splittings for identified pions, kaons, and protons, providing further constraints on their respective relaxation times and the corresponding electromagnetic interaction. These results establish the charge splitting of directed flow as a quantitative probe of the electromagnetic response and the space-time evolution of the QGP.

\end{abstract}

\maketitle

\noindent 1. {\bf Introduction:}
High-energy heavy-ion collisions create a short-lived quark--gluon plasma (QGP), a hot and dense state of strongly interacting matter in which quarks and gluons are deconfined~\cite{Bjorken:1982qr,BRAHMS:2004adc,STAR:2005gfr,PHENIX:2004vcz}. In noncentral collisions, spectator protons generate an intense, rapidly evolving magnetic field~\cite{Skokov:2009qp,Deng:2012pc,Voronyuk:2011jd,Huang:2015oca,Panda:2024ccj,Dash:2023kvr,Panda:2021pvq,Panda:2020zhr,Palni:2024wdy}. The time dependence of this magnetic field induces an electric field through Faraday induction, while the finite electrical conductivity of the QGP gives rise to electric currents that influence the subsequent evolution of both the medium and the electromagnetic (EM) fields~\cite{Gursoy:2014aka,Gursoy:2018yai,Tuchin:2013apa}. The coupled evolution of the EM fields and the QGP is governed by the collision geometry, the electrical conductivity of the medium, and the magnetohydrodynamic response of the QGP~\cite{Deng:2012pc,Gursoy:2014aka,Gursoy:2018yai,Mayer:2024kkv,Nakamura:2022wqr,Panda:2026lbp}. These fields couple to charged constituents of the medium and can generate charge-dependent signatures in final-state observables, including elliptic and directed flow of both light- and heavy-flavor particles~\cite{Gursoy:2014aka,Rybicki:2013qga,Das:2016cwd,Chatterjee:2018lsx,Gursoy:2018yai,STAR:2023jdd,ALICE:2019sgg,Zhang:2022lje,Demircik:2026ctl,PANDA2026140794}.

Experimental measurements of charge-dependent observables at RHIC and the LHC~\cite{STAR:2023jdd,ALICE:2019sgg}, including charge- and strangeness-dependent splittings of produced hadrons at RHIC~\cite{STAR:2023produced}, provide important constraints on the electromagnetic response of the QGP. The charge splitting of directed flow, $\Delta v_1\equiv v_1(h^+)-v_1(h^-)$, exhibits a sign reversal at midrapidity, changing from Au+Au collisions at $\sqrt{s_{\mathrm{NN}}}=200$ GeV for protons to Pb+Pb collisions at $\sqrt{s_{\mathrm{NN}}}=5.02$ TeV for inclusive charged hadrons. This behavior poses a stringent challenge for theoretical descriptions. Effects of electromagnetic field have been investigated with different treatments of the medium response, ranging from calculations without conductivity to approaches incorporating finite conductivity within magnetohydrodynamic frameworks~\cite{Inghirami:2018ziv,Inghirami:2019mkc,Nakamura:2022wqr,Nakamura:2022ssn,Mayer:2024kkv,Gursoy:2018yai,Benoit:2025amn,Rybicki:2013qga,Rybicki:2014yca}, yet a simultaneous quantitative description of the RHIC and LHC measurements, particularly the observed change in the $\Delta v_1$ slope, remains elusive. A quantitative description of $\Delta v_1$ requires a consistent connection between the evolving electromagnetic fields, the QGP response, and the charge-dependent particle distributions at particlization. 

To connect the electromagnetic response of the QGP to final-state observables, previous approaches incorporated spectator-induced fields through drift-velocity corrections to the Cooper--Frye prescription~\cite{Cooper:1974mv,Gursoy:2014aka,Panda:2025lmd,Parida:2025ddt}, while a kinetic theory provides a systematic framework for deriving the corresponding electromagnetic correction to the distribution function~\cite{Panda:2021pvq}. Here, we implement the leading-order electromagnetic correction motivated by the kinetic theory directly in the particlization distribution within CLVisc~\cite{Pang:2018}, without introducing an additional drift velocity. With dynamical electromagnetic fields generated by the conducting QGP, this framework consistently connects the evolving fields, medium response, and freezeout charge distribution underlying the directed-flow splitting.

\begin{figure*}[t]
\includegraphics[width=\textwidth]{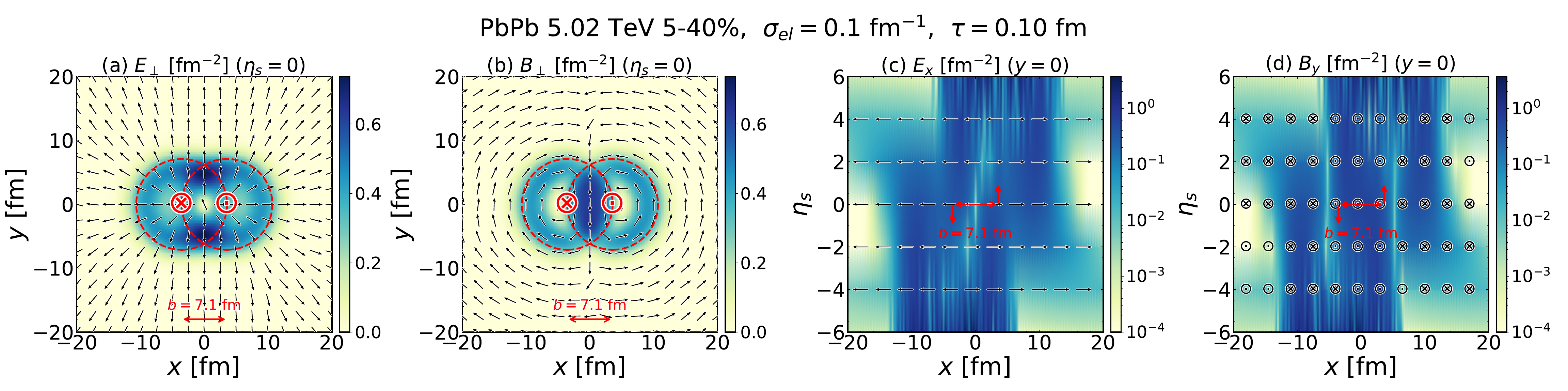}
\caption{Event-averaged laboratory-frame electromagnetic fields at
$\tau=0.10~\mathrm{fm}$, computed from the charged spectator and
participant motion in the same \trento{} geometry.
Panels~(a) and~(b): $E_\perp$ and $B_\perp$ in the transverse plane at
$\eta_s=0$. The color scale gives the magnitude and the arrows give
the direction of the transverse field. The average impact parameter is
$b=7.1~\mathrm{fm}$. Here $\otimes$ denotes motion along
$-\eta_s$ (into the page) and $\odot$ motion along $+\eta_s$ (out of
the page).
Panels~(c) and~(d): $E_x$ and $B_y$ in the $x$--$\eta_s$ plane at
$y=0$. The color scale gives the magnitude. The single-headed arrows
show the direction of the two nuclei along the beam axis
($\pm\eta_s$), and the double-headed arrow marks the impact parameter
$b$. Here $\otimes$ denotes the $+y$ direction (into the page).
\label{fig:eminput}}
\end{figure*}

We first validate the implementation of the electromagnetic response in CLVisc against the semi-analytic solution of resistive magnetohydrodynamics (MHD)~\cite{HaddadiMoghaddam:2020ihi}. We then apply it to Au+Au collisions at $\sqrt{s_{NN}}=200$~GeV and Pb+Pb collisions at $\sqrt{s_{NN}}=5.02$~TeV, using \trento{} initial conditions~\cite{Moreland:2014fqa} and the corresponding experimental kinematic selections. Finally, we calculate the rapidity dependence of the charge splitting of the directed-flow, $\Delta v_1$, and compare with the STAR and ALICE measurements~\cite{STAR:2023jdd,ALICE:2019sgg}. This enables us to disentangle the electric- and magnetic-field contributions and examine their dependence on the 
electrical conductivity and its corresponding relaxation time, thereby establishing how the electromagnetic response of the QGP is reflected in $\Delta v_1$ across RHIC and LHC energies.

\noindent 2. {\bf Model framework:}\label{sec:model}
The QGP evolution in this study is simulated using the $(3+1)$-dimensional viscous hydrodynamic framework CLVisc~\cite{Pang:2018} in Milne coordinates $(\tau,x,y,\eta_s)$~\cite{Bjorken:1982qr}. Electromagnetic effects enter through the Lorentz-force source during the evolution and through a charge-dependent correction to the single particle distribution function at particlization with the first being through:
\begin{equation}
\nabla_\mu T^{\mu\nu}=F^{\nu\lambda}J_\lambda,
\qquad
J^\mu=|q|n_f u^\mu+\sigma_{el}F^{\mu\lambda}u_\lambda ,
\label{eq}
\end{equation}
where $T^{\mu\nu}$, $F^{\mu\nu}$, $J^\mu$, $u^\mu$, $n_f$, and $\sigma_{el}$ denote the energy-momentum tensor, electromagnetic field tensor, electric current, fluid four-velocity, net charge density, and electrical conductivity, respectively. The two terms in $J^\mu$ correspond to the convective and conduction currents. At $\sqrt{s_{\rm NN}}\geq200$~GeV, the net-charge contribution is neglected, leaving the conduction current $J^\mu_{\rm cond}=\sigma_{el}F^{\mu\lambda}u_\lambda$. In our smooth, charge-symmetric setup, this current modifies the common hydrodynamic background but has no direct contribution to $\Delta v_1$; instead, the charge-odd splitting originates from the electromagnetic correction to the particlization distribution.

The electromagnetic implementation is validated against the semi-analytic resistive-MHD solution of Ref.~\cite{HaddadiMoghaddam:2020ihi}, with the validation covering the electromagnetic source, field transformation, and conductivity-dependent response (see Supplemental notes). The electrical conductivity $\sigma_{el}$ controls the induced current and thus the electromagnetic response of the QGP. While $\sigma_{el}$ and the associated relaxation time $\tau_c$ in general may depend on temperature, field strength, momentum, and microscopic properties of charge carriers~\cite{Aarts:2020xtg,Amato:2013oja,Panda:2021pvq,Dwibedi:2025xho}, we nevertheless adopt a constant effective conductivity to isolate the overall response. We relate it to the relaxation time through scaling $\sigma_{el}\propto\tau_c$~\cite{Dwibedi:2025xho} motivated by the kinetic theory, such that increasing $\sigma_{el}$ corresponds to a larger effective $\tau_c$ and a stronger nonequilibrium electromagnetic correction at freeze-out.

At particlization, charged particles acquire an electromagnetic distortion of their momentum distribution away from local equilibrium. We incorporate this effect by adding the leading-order kinetic-theory correction to the equilibrium Cooper--Frye distribution,
$f=f_0+\delta f_{\rm EM}$,
with
\begin{equation}\label{deltafem}
\delta f_{\rm EM}
=
-\tau_c q f_0(1\pm f_0)
\frac{E \cdot p}{T_{\rm frz}\,\left(u \cdot p\right)}.
\end{equation}
where $q$ is the hadron charge, $f_0$ is the equilibrium distribution, $\pm$ accounts for Bose enhancement or Pauli blocking, $T_{\rm frz}$ is the freeze-out temperature, $\tau_c$ is fixed by the conductivity, and $u\cdot p$ is the particle energy in the local fluid rest frame (LRF). The electric field $E^\mu$ is evaluated in the same frame. Since $\delta f_{\rm EM}$ is odd in $q$, oppositely charged hadrons receive opposite momentum distortions.

The electromagnetic fields are calculated from the relativistic motion of charged spectator and participant protons using the same \trento{} geometry for the hydrodynamic evolution~\cite{Deng:2012pc,Gursoy:2014aka,Siddique:2021smf}. They are obtained through a Green's-function formulation~\cite{Gursoy:2014aka,Siddique:2021smf} (with a constant electrical conductivity) in the cylindrical coordinate, transformed to laboratory-frame components, mapped onto the hydrodynamic grid, and extended in space-time rapidity according to the initial longitudinal profile. For each collision system, the fields retain their full space-time dependence and enter both the hydrodynamic source term and the local comoving electric field at particlization. We adopt the convention that the nucleus centered at $+b/2$ moves along $+z$ and that at $-b/2$ along $-z$. The fields are strongest at early times and subsequently evolve as the spectators recede and the conducting medium responds.

Fig.~\Ref{fig:eminput} shows the event-averaged electromagnetic fields for Pb$+$Pb collisions at $\sqrt{s_{NN}}=5.02$~TeV at the initial hydrodynamic time $\tau_0=0.1$~fm$/c$. The corresponding Au$+$Au fields at $\sqrt{s_{\rm NN}}=200$~GeV are generated independently using the appropriate beam energy and collision geometry. For both collision systems, the electromagnetic fields are evaluated from the charged spectator and participant protons within the same event-by-event \trento{} geometry used to initialize the hydrodynamic evolution and are subsequently averaged over 500k events. We consider $(\tau_0,\mathrm{centrality})=(0.6~\mathrm{fm}/c,50$--$80\%)$ for Au$+$Au and $(0.1~\mathrm{fm}/c,5$--$40\%)$ for Pb$+$Pb. 
The viscous-fluid evolution uses $\eta/s=0.16$ and a lattice-QCD equation of state~\cite{Huovinen:2009yb,Borsanyi:2013bia}, followed by Cooper--Frye particlization at $T_{\rm frz}=0.128$~GeV, including resonance decays and the electromagnetic correction $\delta f_{\rm EM}$.


\noindent
\textbf{3. Charge splitting of directed flow:}
The directed-flow coefficient is obtained from the final-state particle distribution as ;
\begin{equation}
v_1(y)=
\frac{
\displaystyle
\int p_T dp_Td\phi
\cos\phi
\frac{d^3N}{p_Tdp_Td\phi dy}
}{
\displaystyle
\int p_T dp_Td\phi
\frac{d^3 N}{p_Tdp_Td\phi dy}
}.
\end{equation}
Here, $y$, $p_T$, and $\phi$ denote the rapidity, transverse momentum, and azimuthal angle for each particle species. We investigate the charge splitting
$\Delta v_1\equiv v_1(h^+)-v_1(h^-)$
arising from the charge-odd electromagnetic correction at particlization.
The spectator nuclei move in opposite directions along the beam axis and are separated by a finite impact parameter, generating a charge-dependent Lorentz force. Under rapidity reflection, the laboratory electric contribution changes sign, while the magnetic contribution inherits the same odd rapidity dependence through its coupling to the longitudinal flow velocity, which is rapidity-odd, with the magnetic field itself being rapidity-even. Consequently, both components of the Lorentz force contribute to the observed rapidity-odd behavior of $\Delta v_1(y)$.



\begin{figure}[ht]
\centering
\includegraphics[width=\linewidth]{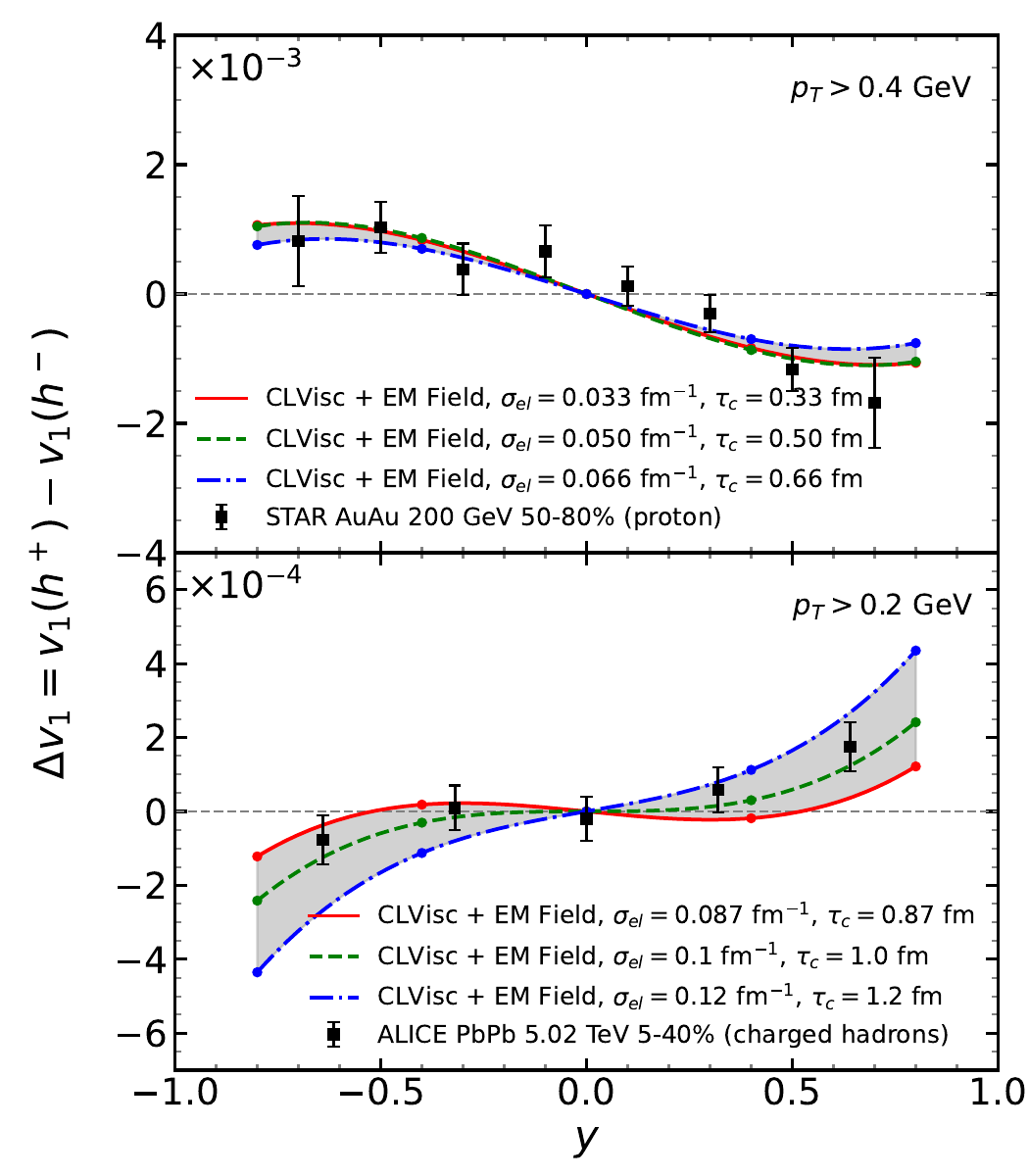}
\caption{
\textit{Upper panel}: Proton charge splitting of directed flow,
$\Delta v_1=v_1(p)-v_1(\bar p)$, as a function of rapidity in
Au$+$Au collisions at $\sqrt{s_{NN}}=200$ GeV for 50--80$\%$ centrality,
compared with STAR data~\cite{STAR:2023jdd}. The calculation uses the
same particle species and kinematic selections as the measurement.
\textit{Lower panel}: Inclusive charged-hadron charge splitting,
$\Delta v_1=v_1(h^+)-v_1(h^-)$, for $\pi+K+p$ as a function of
rapidity in Pb$+$Pb collisions at $\sqrt{s_{NN}}=5.02$ TeV for
5--40$\%$ centrality, compared with ALICE data~\cite{ALICE:2019sgg}.
The particle species and kinematic selections follow the experimental
analysis. The band denotes the theoretical variation obtained by varying
the electrical conductivity around its best-fit value over the
$1\sigma$ range.}
\label{fig:auau-pbpb}
\end{figure}

  Shown in the upper panel of Fig.~\eqref{fig:auau-pbpb} is the charge splitting of directed flow for protons, $\Delta v_1=v_1(p)-v_1(\bar p)$, in Au$+$Au collisions at $\sqrt{s_{\rm NN}}=200$~GeV for $50$--$80\%$ centrality, compared with STAR data. Both show a negative midrapidity slope. The inclusive charged-hadron splitting, $\Delta v_1=v_1(h^+)-v_1(h^-)$, shown in the lower panel for $\pi+K+p$ in Pb$+$Pb collisions at $\sqrt{s_{\rm NN}}=5.02$~TeV and $5$--$40\%$ centrality, has a positive slope at midrapidity, consistent with the ALICE measurement. The electrical conductivity $\sigma_{\rm el}$ is the model-dependent parameter varied within the current framework to describe the experimental data.  Its variation modifies the electromagnetic-field evolution, controls the electromagnetic distortions of momentum distribution at particlization and, consequently, the charge splitting $\Delta v_1$. The error band represents the variation of $\sigma_{\rm el}$ within its $1\sigma$ confidence interval, as determined from the $\chi^2$ analysis. The results are also sensitive to the initial time $\tau_0$, as discussed in the Supplemental notes.
  As discussed above, $\sigma_{\rm el}$ and the relaxation time $\tau_c$ are proportional to each other, so varying the conductivity also changes $\tau_c$. The current experimental data can also be described within the uncertainty of the model using the same $\tau_c$ at the freeze-out however, with different conductivities at RHIC and LHC, within the same partialization prescription.

\textbf{Interplay of Electric and Magnetic Responses:}

\begin{figure}[h]
\includegraphics[width=\columnwidth]{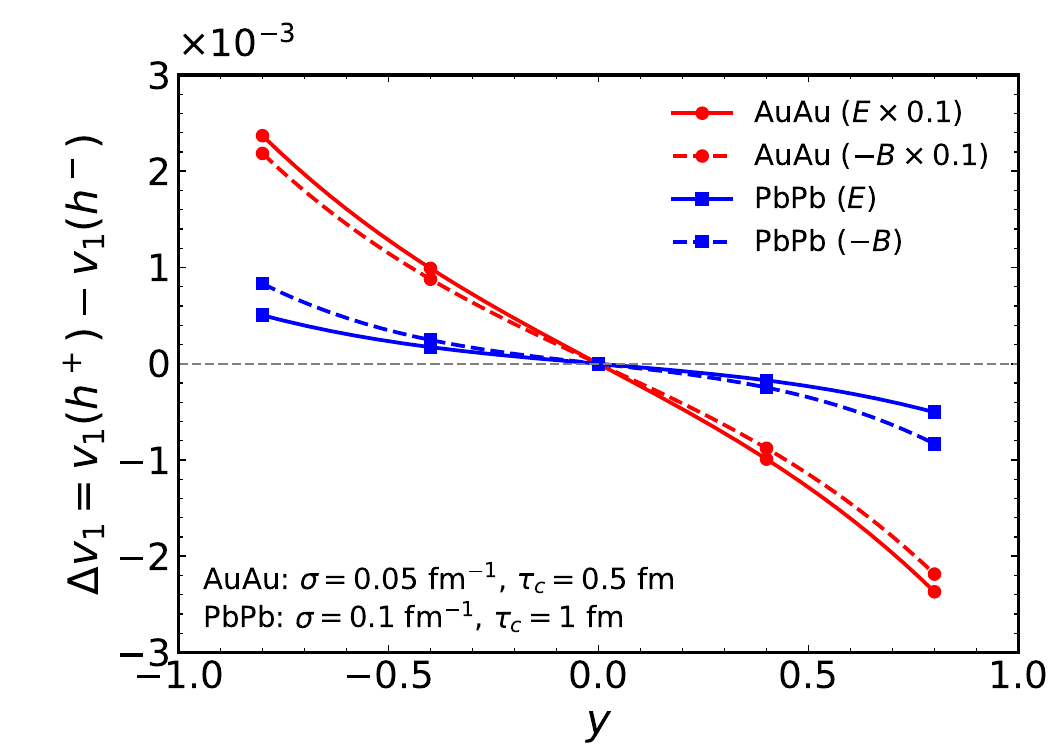}
 \caption{
 Charged-hadron splitting of directed flow,
$\Delta v_1=v_1(h^+)-v_1(h^-)$, as a function of rapidity $y$
for Au$+$Au collisions at $\sqrt{s_{NN}}=200$ GeV (50--80 $\%$ centrality,
pions) and Pb$+$Pb collisions at $\sqrt{s_{NN}}=5.02$ TeV
(5--40 $\%$ centrality, pions).  The electric- and magnetic-field contributions are shown separately; the magnetic-field contributions are multiplied by $-1$ so that their magnitudes can be compared directly with the electric ones. The Au$+$Au proton results are multiplied by 0.1 for easy comparison.
}
\label{fig:auau-proton-pbpb-pion}
\end{figure}

\noindent
The origin of the sign change in $\Delta v_1$ lies in the momentum-dependent electromagnetic distortion of the phase-space distribution function at particlization, derived from the relativistic Boltzmann equation as given in Eq.~\eqref{deltafem},
where $E_\mu = F_{\mu\nu}u^\nu$ is the electric field evaluated in the LRF. Taking representative fields such as $E_x$ and $B_y$ in lab frame, the LRF electric field in $\delta f_{\rm EM}$ is  $F_{x\nu} u^\nu \sim \gamma(E_x - v_{z}  B_y)$, where $\gamma = 1/\sqrt{1- v_{z}^{2}}$. This structure reveals that the Lorentz force driving the distortion of the phase-space distribution contains an electric and a velocity-coupled magnetic contribution with opposite signs, leading to their competing contributions to the charge splitting of the directed flow $\Delta v_1$ with opposite signs as shown in Fig.~\eqref{fig:auau-proton-pbpb-pion}.

Because the electric and magnetic fields possess fundamentally distinct space-time profiles, their momentum-integrated contributions to the splitting have different dependencies on rapidity and  the conductivity (and thus the relaxation time). As illustrated in Fig.~\eqref{fig:auau-proton-pbpb-pion} for pions at both RHIC and LHC energies, the electric contribution drives a negative midrapidity slope, whereas the magnetic contribution favors a positive slope. At RHIC energies with the conductivity $\sigma_{el}=0.05\ \rm{fm}^{-1}$, the magnitude of the electric contribution after the space-time integral is larger than the magnetic contribution, leading to
a negative midrapidity slope of the net contribution. An enhanced conductivity $\sigma_{el}$ at LHC energies, on the other hand,  extends the lifetime of the magnetic field on the freeze-out hypersurface and makes the magnitude of the magnetic contribution larger than the electric contribution. The net contribution in this case has a positive midrapidity slope.  While the relaxation time $\tau_c$ sets the overall magnitude of the splitting, the competing electric and magnetic fields dictate its rapidity dependence and sign of the slope. Notably, the magnitude of the charge splitting is larger at lower collision energies than at higher energies, where the rapidity slope also changes sign, consistent with experimental observations~\cite{STAR:2023jdd}.


\begin{figure}[h]
\centering
\includegraphics[width=\linewidth]{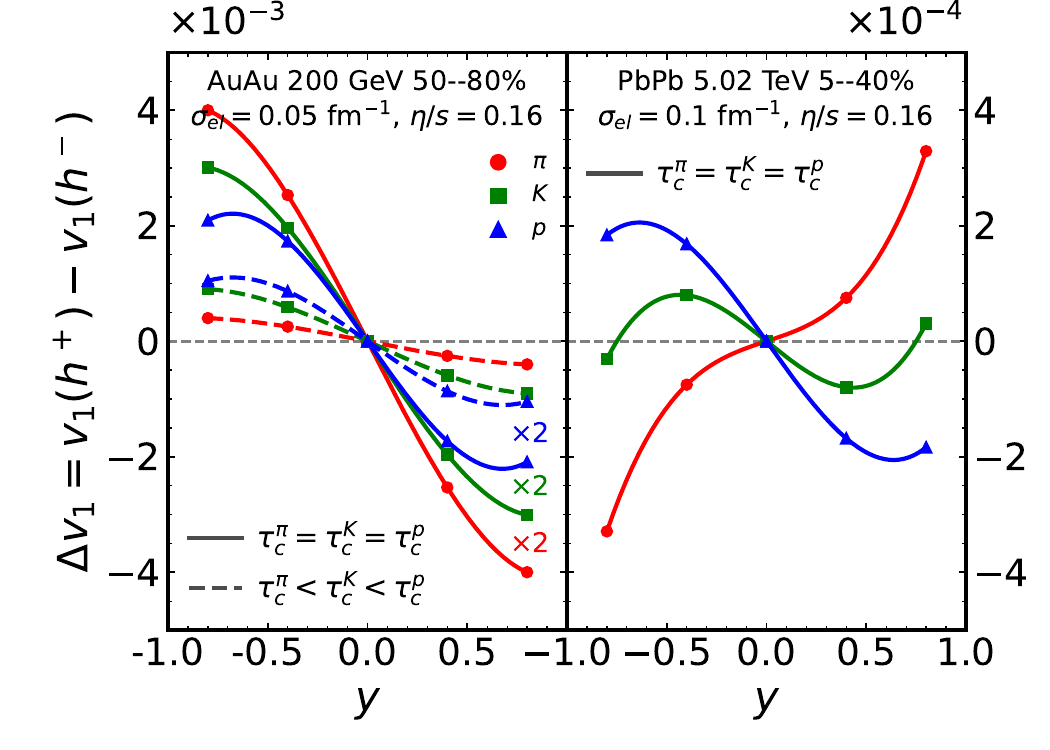}
\caption{
Predictions for the charge-odd directed flow,
$\Delta v_1=v_1(h^+)-v_1(h^-)$, of pions, kaons, and protons. \textit{Left panel}:
Au$+$Au collisions at $\sqrt{s_{\rm NN}}=200$ GeV for 50--80 $\%$
centrality with $\sigma_{el}=0.05$ fm$^{-1}$. \textit{Right panel}: Pb$+$Pb collisions at
$\sqrt{s_{\rm NN}}=5.02$ TeV for 5--40 $\%$ centrality with
$\sigma_{el}=0.1$ fm$^{-1}$. The results illustrate the sensitivity of the
identified-particle splitting to the assumed relaxation-time hierarchy.
For Au$+$Au, both species-independent and species-dependent relaxation
times are shown. The Pb$+$Pb results illustrate the corresponding
species dependence in the regime where the electromagnetic momentum
projection is dominated more strongly by the field contribution.}
\label{fig:species}
\end{figure}

\textbf{Mass hierarchy of the charged splitting $\Delta v_1$:}
With the adjusted colliding energy dependence of the conductivity (relaxation time), our model can successfully describe the charge splitting of the directed flow (both the magnitude and the sign of the rapidity slope) for protons in Au$+$Au and inclusive charged hadrons in Pb$+$Pb using a consistent electromagnetic and hydrodynamic framework. The same framework can also predict the mass dependence as shown in Fig.~\eqref{fig:species} where predictions for the charged directed flow splitting of pions, kaons, and protons in Au$+$Au collisions at $\sqrt{s_{\rm NN}} = 200$~GeV (left panel) and Pb$+$Pb collisions at $\sqrt{s_{\rm NN}} = 5.02$~TeV (right panel) are presented. For Au$+$Au, results are shown for both mass-independent and mass-hierarchical relaxation times. These mass-dependent responses originate from the interplay of the thermal distribution function, particle mass, momentum projection $E\cdot p$, and the relaxation time $\tau_c$ entering $\delta f_{\rm EM}$. For a common $\tau_c$, thermal and kinematic weights generate distinct mass hierarchies: at RHIC energies, the mass dependence of the thermal distribution is prominent, favoring larger splitting for lighter hadrons, whereas at LHC energies, harder momentum spectra give greater weight to the $E\cdot p$ projection, modifying this simple mass ordering. These findings yield clear predictions for the electromagnetic contribution to $\Delta v_1$  of identified hadrons, with pions exhibiting the strongest sensitivity to electromagnetic fields and driving the charged splitting at LHC energies.

With the mass dependence in mind, introducing a mass hierarchy in the relaxation time (a heavier particle
has a larger $\tau_c$) can therefore significantly modify their electromagnetic response. Because $\delta f_{\rm EM} \propto \tau_c E\cdot p$, the larger relaxation times of kaons and protons can compensate for the mass-dependent thermal suppression, allowing the proton $\Delta v_1$ to become comparably enhanced when this mass dependence of the relaxation time is sufficiently strong. 

While the absolute magnitudes depend on the specific $\tau_c$ prescription, the resulting mass ordering directly reflects the interplay between electromagnetic fields and microscopic relaxation dynamics. Ultimately, a precision measurement of these mass-dependent charge splittings offers a sensitive probe of the underlying transport properties of the QGP.

\noindent
\textbf{4. Summary and conclusions:}
We investigate the charge splitting of the directed flow $\Delta v_1$ generated by the electromagnetic correction to the particlization distribution in relativistic heavy-ion collisions. Within a unified electromagnetic and hydrodynamic framework, a beam-energy dependent conductivity (or relaxation time) can reproduce the magnitude and a negative midrapidity slope of the proton $\Delta v_1$ in Au$+$Au collisions at RHIC energies, as well as the positive slope of inclusive charged hadrons in Pb$+$Pb collisions at LHC energies.

The underlying mechanism is governed by the competition between the electric and magnetic contributions to the momentum-dependent correction $\delta f_{\rm EM}$ given in Eq.~\eqref{deltafem}.
A hypersurface-weighted projection of this correction shows that the electric contribution precedes the magnetic fields at RHIC, whereas the magnetic contribution becomes more important at the LHC for a beam-energy dependent conductivity (relaxation time). Since the two contributions enter with opposite signs, their changing relative strength naturally explains the reversal of the $\Delta v_1$ midrapidity slope between RHIC and LHC energies. Note that we used constant ($\sigma_{\rm el}$, $\tau_c$) in our current study which should be considered as the average value throughout the evolution and freeze-out at each beam energy. In principle, these transport parameters should depend on local temperature and other local variables and evolve with time.  This should be considered in future improvement of the framework. 

The current framework however also predicts a mass-dependent charge splitting $\Delta v_1$  for pions, kaons and protons.
When the $\tau_c E\cdot p$ contribution becomes sufficiently important, a mass hierarchy of the relaxation times can modify or even reverse the conventional mass ordering. Measurements of $\Delta v_1$
of identified hadrons therefore provide additional sensitivity to the relaxation dynamics of the electromagnetic response in the particlization process.

Overall, charge-dependent directed flow, together with its beam-energy and mass dependence, provides a sensitive probe of the electromagnetic response and transport properties of the QGP.


\noindent{\bf Acknowledgments:} We thank Shu-Su Shi for helpful discussions.
This work was supported in part by the NSFC under grant Nos.~12435009 and 12535010 and the Outstanding Leading Talent Team Program of Central China Normal University (XJ2026000302). Numerical calculations were carried out at NSC3 at CCNU.


\bibliographystyle{apsrev4-2}
\bibliography{reference}

\appendix
\newpage
\section*{Supplemental materials}\label{supp}

\noindent {\bf 1: Electromagnetic source term in Milne coordinates:}\label{app:source}
The electromagnetic contribution to the hydrodynamic evolution is obtained
from the covariant conservation of the matter ($T_{\rm m}^{\mu\nu}$) and electromagnetic
energy--momentum tensors ($T_{\rm f}^{\mu\nu}$). Writing the total we have
\begin{equation}
T^{\mu\nu}=T_{\rm m}^{\mu\nu}+T_{\rm f}^{\mu\nu},
\qquad
T_{\rm f}^{\mu\nu}
=
F^{\mu\lambda}F^\nu_{\ \lambda}
-\frac{1}{4}g^{\mu\nu}F_{\lambda\kappa}F^{\lambda\kappa},
\end{equation}
and using Maxwell's equations,
\begin{equation}
\nabla_\mu F^{\mu\nu}=J^\nu,
\qquad
\nabla_\mu\widetilde{F}^{\mu\nu}=0,
\end{equation}
the matter sector satisfies
\begin{equation}
\nabla_\mu T_{\rm m}^{\mu\nu}
=
F^{\nu\lambda}J_\lambda .
\label{eq:app_source}
\end{equation}

For the systems considered in this work, the net electric charge density
is neglected and the current is given by the conduction contribution,
$J^\mu
=
\sigma_{el} F^{\mu\lambda}u_\lambda $.
The corresponding electromagnetic source term can therefore be written as:
\begin{equation}
s^\nu
=
\sigma_{el} F^{\nu\lambda}F_{\lambda\mu}u^\mu ,
\label{eq:app_force}
\end{equation}
which is evaluated locally on the hydrodynamic grid and included in the
CLVisc evolution equations.

The electromagnetic fields are calculated in the laboratory Cartesian
frame and subsequently transformed to Milne coordinates
$x^\mu=(\tau,x,y,\eta_s)$. With
\begin{equation}\nonumber
t=\tau\cosh\eta_s,
\qquad
z=\tau\sinh\eta_s,
\end{equation}
the field-strength tensor in Milne coordinates is obtained from
\begin{equation}
F^{\mu\nu}
=
\frac{\partial x^\mu}{\partial \bar{x}^{\alpha}}
\frac{\partial x^\nu}{\partial \bar{x}^{\beta}}
\bar{F}^{\alpha\beta},
\label{eq:app_transform}
\end{equation}
where $\bar{x}^\alpha=(t,x,y,z)$ denotes the laboratory Cartesian
coordinates. The transformed fields are then combined with the local
fluid four-velocity to construct the source term in Eq.~(\ref{eq:app_force}).


\noindent {\bf 2: Relaxation-time derivation of the electromagnetic correction:}\label{app:deltaf}
The electromagnetic correction to the particlization distribution follows
from the relativistic Boltzmann equation in the relaxation-time
approximation,
\begin{equation}
p^\mu\partial_\mu f
+
qF^{\mu\nu}p_\nu
\frac{\partial f}{\partial p^\mu}
=
-\frac{p\cdot u}{\tau_c}(f-f_0),
\label{eq:app_boltzmann}
\end{equation}
where
\begin{equation}
f_0
=
\left[\exp(\beta u\cdot p - \beta \mu)\mp1\right]^{-1}.
\end{equation}
Writing $f=f_0+\delta f_{\rm EM}$ and retaining terms linear in the electromagnetic
field, the Lorentz-force term acting on the equilibrium distribution gives
\begin{equation}
qF^{\mu\nu}p_\nu
\frac{\partial f_0}{\partial p^\mu}
=
-q\beta f_0(1\pm f_0)\,
(E\cdot p),
\end{equation}
where the electric field is evaluated in the local fluid rest frame,
$E^\mu=F^{\mu\nu}u_\nu$.
Balancing this source term against the relaxation term yields
\begin{equation}
\delta f_{\rm EM}
=
-\tau_c q f_0(1\pm f_0)
\frac{E\cdot p}
{T_{\rm frz}\,u\cdot p},
\label{eq:app_deltaf}
\end{equation}
which is the correction used in the Cooper--Frye particlization.

The explicit factor of the particle charge makes this correction
charge-odd and therefore provides the leading electromagnetic contribution
to the charge splitting of directed flow considered in this work. The
relaxation time entering Eq.~(\ref{eq:app_deltaf}) is not treated as an
independent parameter; it is determined from the same conductivity used in the electromagnetic evolution, as described in the main text.

\noindent {\bf 3: Validation against a semi-analytic solution:}\label{app:validation}
We validate the implementation of the electromagnetic source term by
comparing the CLVisc evolution with the semi-analytic solution of
relativistic resistive magnetohydrodynamics presented in
Ref.~\cite{HaddadiMoghaddam:2020ihi}. The benchmark considers a
$(1+1)$-dimensional longitudinally expanding system for which the
electromagnetic fields and the corresponding hydrodynamic response are
known analytically.
\begin{figure}[h]
\includegraphics[width=\columnwidth]{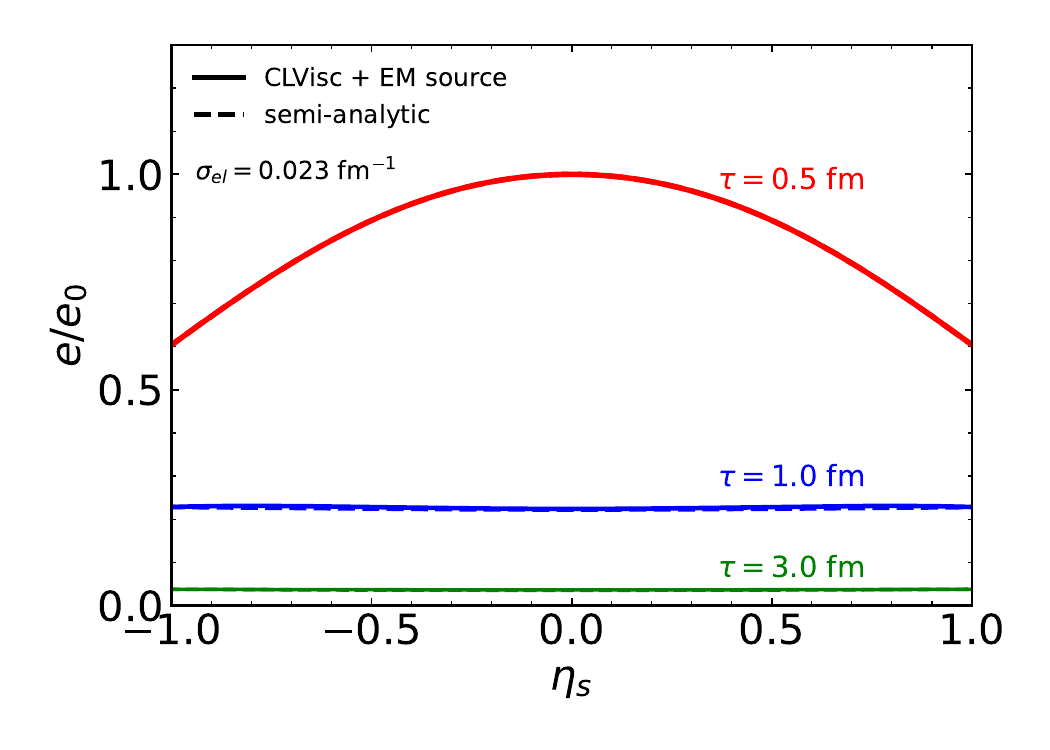}
\caption{Energy-density ratio $e/e_0$ as a function of space-time rapidity
$\eta_s$ at $\tau=0.5$, 1, and $3$ fm/$c$ for
$\sigma_{el}=0.023$ fm$^{-1}$. Solid lines show the CLVisc results, while dashed
lines denote the semi-analytic solution of Ref.~\cite{HaddadiMoghaddam:2020ihi}.
The agreement verifies the implementation of the electromagnetic source
term in the hydrodynamic evolution.}
\label{fig:ed}
\end{figure}
For the comparison, the electromagnetic fields specified by the
semi-analytic solution are supplied to CLVisc in the laboratory frame and
transformed internally to Milne coordinates according to
Eq.~(\ref{eq:app_transform}).

\begin{figure}[h]
\includegraphics[width=\columnwidth]{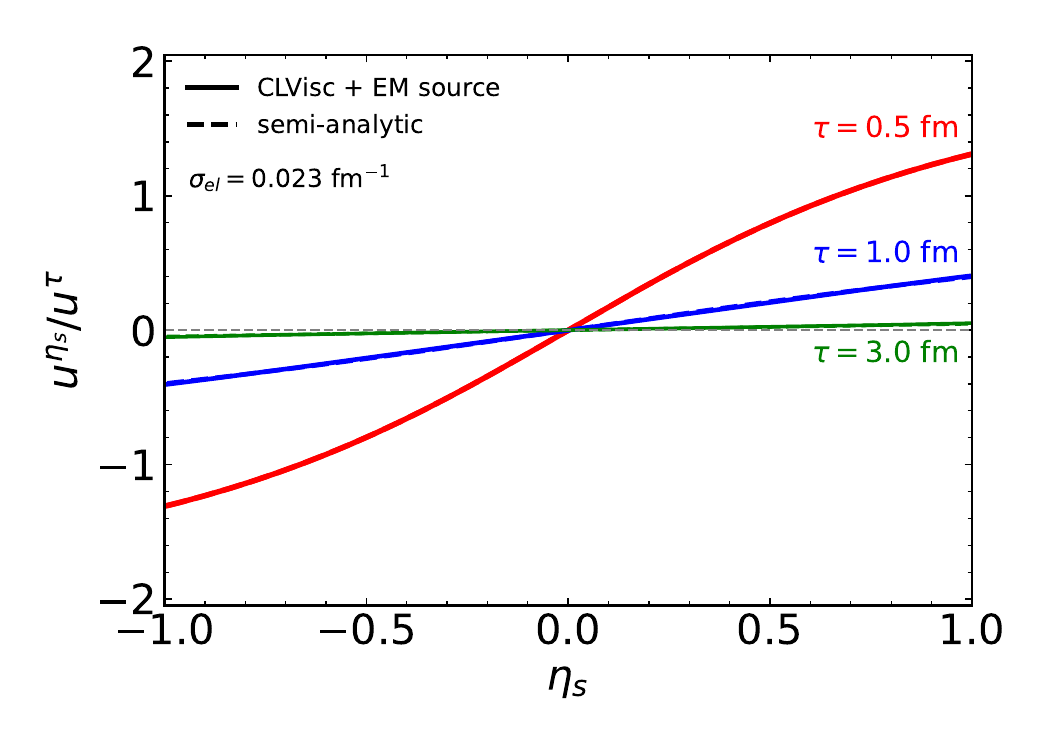}
\caption{Longitudinal flow $u^{\eta_s}/u^\tau$ as a function of
$\eta_s$ at $\tau=0.5$, $1$, and $3$ fm/$c$ for
$\sigma_{el}=0.023$ fm$^{-1}$. Solid lines show the CLVisc results and dashed
lines the semi-analytic solution of Ref.~\cite{HaddadiMoghaddam:2020ihi}.
The antisymmetric rapidity dependence and its proper-time evolution are
reproduced quantitatively.}
\label{fig:veta}
\end{figure}

\begin{figure}
\includegraphics[width=\columnwidth]{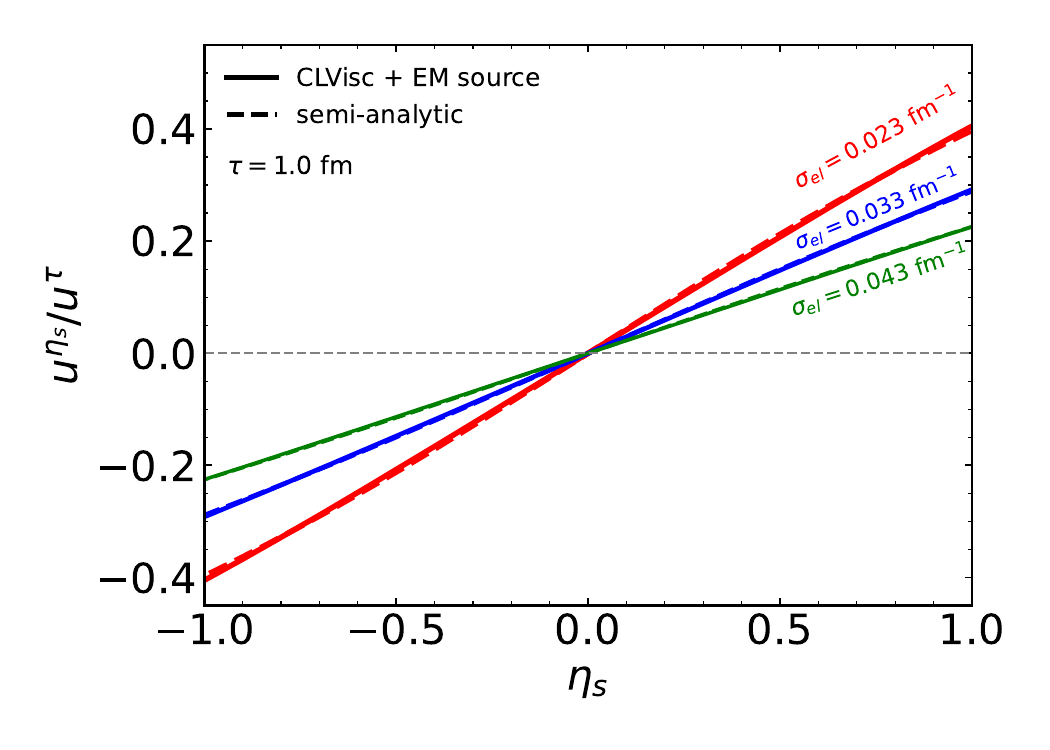}
\caption{Conductivity dependence of the longitudinal flow
$u^{\eta_s}/u^\tau$ as a function of $\eta_s$ at $\tau=1$ fm/$c$ for
$\sigma_{el}=0.023$, $0.033$, and $0.043$ fm$^{-1}$. Solid lines show the
CLVisc results and dashed lines the corresponding semi-analytic solutions
of Ref.~\cite{HaddadiMoghaddam:2020ihi}. The numerical evolution reproduces
the conductivity dependence of the analytic solution and approaches the
magnetized limit with increasing conductivity.}
\label{fig:sigma}
\end{figure}

The same conductivity is used in the
hydrodynamic source term as in the benchmark solution. This provides a
direct test of the field transformation, the Lorentz-force source term,
and its coupling to the hydrodynamic evolution.

Figure~\eqref{fig:ed} shows the energy-density ratio $e/e_0$ as a function of
space-time rapidity at several proper times. The CLVisc results reproduce
the semi-analytic solution over the full rapidity range considered,
including the characteristic initial profile and its subsequent evolution.
Figure~\eqref{fig:veta} compares the longitudinal flow,
$u^{\eta_s}/u^\tau$, and shows agreement with the analytic result, including
its rapidity-odd dependence and its decrease with proper time.
The conductivity dependence is further examined in
Fig.~\eqref{fig:sigma}. The numerical results follow the analytic behavior
over the range of conductivities considered and approach the expected
magnetized limit with increasing conductivity.
These comparisons demonstrate that the electromagnetic source term,
coordinate transformation, conductivity dependence, and unit conventions
are implemented consistently. The quantitative agreement with the
semi-analytic benchmark provides a validation of the electromagnetic
hydrodynamic module used for the charge-dependent directed-flow
calculations presented in the main text.

\noindent {\bf 4: Sensitivity of $\Delta v_1$ to the initial time $\tau_0$:}\label{app:sen-tau0}

\begin{figure}[h]
\includegraphics[width=\columnwidth]{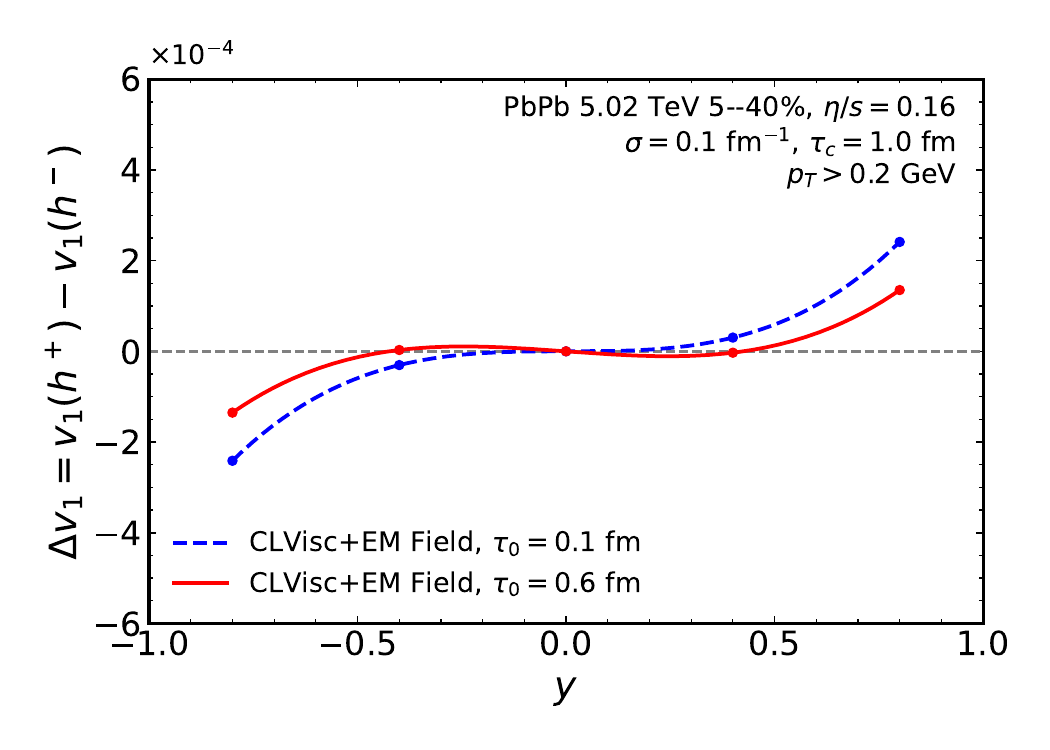}
\caption{Rapidity dependence of the charged-hadron directed-flow splitting, $\Delta v_1 = v_1(h^+) - v_1(h^-)$, in Pb$+$Pb collisions at $\sqrt{s_{NN}} = 5.02$~TeV, computed with viscous hydrodynamics including the electromagnetic-field source term and the $\delta f_{\rm EM}$ correction at freeze-out. The two curves compare hydrodynamic initialization time at $\tau_0 = 0.1$~fm (dashed) and $\tau_0 = 0.6$~fm (solid), illustrating its sensitivity to $\Delta v_1$.}\label{fig:sen-tau0}
\end{figure}

\noindent
We examine the sensitivity of $\Delta v_1$ to the hydrodynamic initialization time $\tau_0$. Because the electromagnetic fields are strongest at early times, delaying the initialization reduces their influence on the subsequent evolution. For illustration, we increase $\tau_0$ from $0.1$ to $0.6$~fm/$c$ in Pb$+$Pb collisions at $\sqrt{s_{NN}}=5.02$~TeV. As shown in Fig.~\eqref{fig:sen-tau0}, the delayed initialization flattens the rapidity dependence of $\Delta v_1$, reducing the magnitude of its slope. This dependence underscores the sensitivity of the charge splitting to the early-time electromagnetic response.

\end{document}